\documentclass[12pt]{article}
\usepackage{epsfig,amsfonts,amssymb}
\usepackage{hyperref}
\usepackage{cite}
\input epsf.sty
\usepackage{cite}

\def\ZZZ{{\hbox{ \kern -1.6mm Z\kern-1.6mm Z}}}
\def\RRR{{\hbox{ R\kern-2.4mm R}}}
\def\LLL{{\hbox{ L\kern-2.0mm L}}}
\def\CCC{{\hbox{ C\kern-2.0mm C}}}
\def\zzz{{\hbox{z\kern-1mm z}}}

\newcommand{\vt}{\vartheta}

\newcommand{\qeq}{{\hbox{=\kern-2.3mm ? \kern.5mm }}}
\renewcommand{\qeq}{=}

\newcommand{\wt}{\widetilde}
\newcommand{\wh}{\widehat}

\newcommand{\be}{\begin{equation}}
\newcommand{\ee}{\end{equation}}
\newcommand{\ben}{\begin{eqnarray}\displaystyle}
\newcommand{\een}{\end{eqnarray}}

\newcommand{\bea}[1]{\begin{eqnarray}\label{#1} }
\newcommand{\eea}{\end{eqnarray}}

\newcommand{\refb}[1]{(\ref{#1})}

\def\one{{\hbox{ 1\kern-.8mm l}}}
\def\zero{{\hbox{ 0\kern-1.5mm 0}}}

\begin{document}

\baselineskip 24pt

\begin{center}
{\Large \bf
U-duality Invariant Dyon Spectrum in type II on $T^6$}

\end{center}

\vskip .6cm
\medskip

\vspace*{4.0ex}

\baselineskip=18pt

\centerline{\large \rm   Ashoke Sen }

\vspace*{4.0ex}

\centerline{\large \it Harish-Chandra Research Institute}

\centerline{\large \it  Chhatnag Road, Jhusi,
Allahabad 211019, INDIA}

\vspace*{1.0ex}
\centerline{E-mail:  sen@mri.ernet.in, ashokesen1999@gmail.com}

\vspace*{5.0ex}

\centerline{\bf Abstract} \bigskip

We give a manifestly U-duality invariant formula for the
degeneracy of 1/8 BPS dyons in type II string theory on $T^6$
for a U-duality invariant subset of charge vectors. Besides
depending on the Cremmer-Julia invariant the degeneracy also
depends on other discrete invariants of $E_{7(7)}(\ZZZ)$.

\vfill \eject

\baselineskip=18pt

Type IIA string theory compactified on $T^6$ is known to have
an $E_{7(7)}(\ZZZ)$ U-duality invariance\cite{9410167}. 
Thus we would expect
the spectrum of 1/8 BPS dyons in this theory to be invariant under
this symmetry. {\it A priori} this would require us 
to make a U-duality
transformation not only on the charges but also on the moduli.
One can partially avoid this problem by using the index measured by
the helicity trace $B_{14}$\cite{9708062,9708130}. 
This is known to not change under
continuous variation of the moduli. However it can in principle jump
across walls of marginal stability on which the 1/8 BPS dyon
may decay into a pair of half-BPS dyons. It was however shown
in \cite{0803.1014} that such
walls of marginal stability are absent for states which 
have a positive value of the 
Cremmer-Julia invariant\cite{cremmer,9602014}, -- the unique
quartic combination of the charges which is invariant under the
continuous $E_{7(7)}$ transformation.
Thus for such states one
should be able to express the index $B_{14}$ as a function of the
charges alone, and furthermore this formula must be invariant under
the discrete U-duality group $E_{7(7)}(\ZZZ)$ acting on the
charges.

For a special class of charge vectors a formula for the index
$B_{14}$ has been derived in 
\cite{0506151,0506228,0803.1014}. 
In principle we can
use
the U-duality invariance of the theory to extend the result to all other
charge vectors which are U-dual to the ones analyzed earlier.
Our goal in this paper is to express the formula for $B_{14}$ as a
manifestly U-duality invariant function of the charges. During
this analysis we shall also find that the U-duality orbit of the
charge vectors analyzed earlier do not cover the full set of
allowed charges in the theory, and we shall be able to express this
restriction in a manifestly U-duality invariant manner.

We begin by introducing some notations and conventions. Type
IIA string theory compactified on $T^6$ has 28 U(1) gauge fields
of which 12 arise in the NSNS sector and the rest arise in the
RR sector. Thus a generic state is characterized by 28 electric and
28 magnetic charges which together transform in the {\bf 56}
 representation of the U-duality 
 group $E_{7(7)}(\ZZZ)$\cite{9410167}.
 We shall denote this 56 dimensional charge vector by $q_a$ 
 ($1\le a\le 56$).  $q_a$'s will be 
 normalized so that $q_a\in\ZZZ$ and
 transform into integer linear combinations of each other under
 $E_{7(7)}(\ZZZ)$.
 It is often useful to examine the transformation
 properties of the charges under a special subgroup of the
 U-duality group containing the T-duality group $SO(6,6;\ZZZ)$
 and the electric-magnetic S-duality group $SL(2,\ZZZ)$. Under
 the $SL(2,\ZZZ)\times SO(6,6;\ZZZ)$ subgroup the charge vector
 transforms in the ${\bf (2,12)\oplus (1,32)}$ representation. Of
 these the {\bf (2,12)} component can be identified as the
 NSNS charges and the ${\bf (1,32)}$ component
 can be identified as the
 RR charges. We shall denote the {\bf (2,12)} part of $q_a$ as
 $M_{i\alpha }$ with $1\le i\le 12$, $1\le\alpha\le 2$ and the
 ${\bf (1,32)}$ component of $q_a$ as $N_s$ ($1\le s\le 32$).
We can identify the components $M_{i1}$ as the electric charges
$Q_i$ in the NSNS sector and the components $M_{i2}$ as the
magnetic charges $P_i$ in the NSNS sector. 

Let us now note some important facts about $E_{7(7)}(\ZZZ)$
representations\cite{slansky}:

\begin{enumerate}

\item
If we have two  vectors
$q_a$ and $q_a'$ in the {\bf 56} 
representation then there is a unique
anti-symmetric
bilinear $\LLL_{ab} q_a q'_b$ which is a singlet of the
continuous 
$E_{7(7)}$ group.
We shall normalize $\LLL_{ab}$ such that the part of 
$\LLL_{ab} q_a q'_b$ involving the NSNS charges take the
form $\epsilon_{\alpha\beta} L_{ij} M_{i\alpha} M'_{j\beta}$,
where $\epsilon_{\alpha\beta}$ is a totally antisymmetric
tensor with the $\epsilon_{12}=-\epsilon_{21}=1$ and $L_{ij}$
is the $O(6,6)$ invariant metric with 6 eigenvalues 1 and 6
eigenvalues $-1$.

\item
The bilinear $q_a q_b$ can be decomposed into a
{\bf 133} and a {\bf 1463} representation of 
$E_{7(7)}$. Let us focus on the
{\bf 133} component of this bilinear.
After suitable normalization one can ensure that all the 
133 elements
are integers and that they transform into integer linear
combinations of each other under an $E_{7(7)}(\ZZZ)$
transformation. We denote by $\psi(q)$ the gcd of all these
133 elements:
\be \label{e0}
\psi(q) \equiv \gcd (q\otimes q)_{\bf 133}\, .
\ee
Then
$\psi(q)$ must remain invariant under an 
$E_{7(7)}(\ZZZ)$ transformation. 

\item
The trilinear
$q_a q_b q_c$ decomposes into {\bf 56},  {\bf 6480} and
{\bf 24320} representations of $E_{7(7)}$. Let us focus
on the {\bf 56} part and denote these 56 components by
$\wt q_a$:
\be \label{e0a}
\wt q \equiv (q\otimes q\otimes q)_{\bf 56}\, .
\ee 
The normalization of $\wt q$ is fixed as follows.
We can construct an $E_{7(7)}(\ZZZ)$ invariant
$\Delta(q)$ as
\be \label{e2}
\Delta(q) = {1\over 2} \LLL_{ab}\wt q_a   q_b\, .
\ee
$\Delta(q)$ is the well known Cremmer-Julia invariant.
We shall normalize $\wt q_a$ such that $\Delta(q)$ has
the following normalization 
\be \label{e3}
\Delta(q)=Q^2 P^2 - (Q\cdot P)^2 +\cdots 
\ee
where the inner products of $Q$ and $P$ are calculated using
the $SO(6,6)$ invariant metric $L_{ij}$ and
$\cdots$ denotes terms involving RR charges. 
With this normalization of $\wt q_a$, we define
\be \label{e4}
\chi(q) \equiv \gcd(q\wedge \wt q) = \gcd\{q_a \wt q_b-q_b\wt q_a\}\, .
\ee
Since
$\wt q_a$ transforms in the same way as $q_a$, the components
of $\wt q_a$ transform into integer linear combinations of
each other under $E_{7(7)}(\ZZZ)$. It   follows from this that 
$\chi(q)$
is invariant under an $E_{7(7)}(\ZZZ)$ transformation. 

\end{enumerate}

We are now ready to state our result. First we shall state the
restriction on the charges for which our result is valid. This
restriction can be stated in two parts:
\begin{enumerate}
\item The charge vector $q$ must be such that its {\bf (1,32)} part
can be removed by an $E_{7(7)}(\ZZZ)$ transformation. In other
words it must be U-dual to a configuration where the D-brane
charges vanish.\footnote{We do not know if this condition is trivial,
\i.e.\ whether it holds for all charge vectors for which 1/8 BPS dyon
exists.}

\item We constrain our charge vector
$q$ such that 
\be \label{e1}
\psi(q)=1\, . 
\ee
\end{enumerate}
Given a charge vector satisfying these two conditions, our
result for the dyon spectrum can be expressed
in terms
of $\Delta(q)$ and $\chi(q)$ as follows.  
The index $d(q)\equiv (-1)^{\Delta(q)}B_{14}
= (-1)^{Q\cdot P} B_{14}$ associated
with the charge vector $q$ is given by
\be \label{e5}
d(q)=\sum_{s\in\zzz, 2s|\chi(q)} s\, \wh c(\Delta(q)/s^2)\, ,
\ee
where $\wh c(u)$ is defined through the 
relations\cite{9903163,0506151}
\be \label{ek6.5}
-\vt_1(z|\tau)^2 \, \eta(\tau)^{-6} \equiv \sum_{k,l} \wh c(4k-l^2)\, 
e^{2\pi i (k\tau+l z)}\, .
\ee
$\vt_1(z|\tau)$ and $\eta(\tau)$ are respectively the odd Jacobi
theta function and the Dedekind eta function. 
Since $\chi(q)$ and $\Delta(q)$ are $E_{7(7)}(\ZZZ)$ invariant, the
above formula is manifestly $E_{7(7)}(\ZZZ)$ invariant.

We shall now give the proof of \refb{e5}. Since we have restricted
our charge vector $q$ such that it can be rotated into purely
NSNS charges, and since \refb{e5} is manifestly U-duality
invariant, it is enough to prove \refb{e5} for charges
belonging to the NSNS sector. In this case with the help of
T-duality transformations we can bring the electric and magnetic
components $Q$ and $P$ into a four dimensional subspace,
represented by fundamental string winding and momentum
along two circles $S^1$ and $\wh S^1$ and Kaluza-Klein (KK)
monopole charges and H-monopole charges along the same
circles\cite{wall}. 
Using further S- and T-duality transformations
we can bring the charges into the 
form\cite{0712.0043,0801.0149}
\be \label{e6}
Q=\pmatrix{0 \cr n\cr 0\cr 1}, 
\quad P=\pmatrix{Q_1\cr J\cr Q_5\cr 0}\, ,
\quad n,Q_1,Q_5,J\in\ZZZ, \quad Q_5|J, Q_1\, .
\ee
In this subspace the T-duality invariant metric $L$ takes
the form
\be \label{e7}
L =\pmatrix{ 0 & I_2\cr I_2 & 0}\, ,
\ee
so that we have
\be \label{e8}
Q^2 = 2n, \quad P^2 = 2 Q_1Q_5, \quad Q\cdot P = J\, .
\ee
Also since $Q_5|Q_1,J$, the torsion 
associated with the pair $(Q,P)$
is\cite{0702150}
\be \label{egcd}
\gcd(Q\wedge P)=Q_5\, .
\ee 
In a suitable duality frame and up to signs,
we can interprete $Q$ as representing $n$ units of fundamental
string winding charge and 1 unit of momentum along $S^1$ and
$P$ as representing $Q_1$ units of NS 5-brane wrapped on
$T^4\times S^1$, $J$ units of KK monopole charge associated
with $S^1$ and $Q_5$ units of KK monopole charge associated
with $\wh S^1$. Here $T^4$ denotes the four torus along
directions other than those labelled by $S^1$ and $\wh S^1$.

Let us now investigate the meaning of the constraint \refb{e1}
on $(Q,P)$. $\psi(q)$ is the gcd of the components of the {\bf 133}
representation in the bilinear $q_a q_b$. Now under 
$SL(2,\ZZZ)\times O(6,6;\ZZZ)$ the {\bf 133} representation of
$E_{7(7)}(\ZZZ)$ decomposes into {\bf (3,1) +(1,66) + (2,32$'$)}.
If the original $q_a$'s have vanishing components corresponding
to the RR directions then the ${\bf (2,32')}$ component vanishes
and the {\bf (3,1)} and {\bf (1,66)} components correspond to,
respectively
\be \label{e8a}
\{ Q^2/2, P^2/2, Q\cdot P\} \quad \hbox{and} \quad
\{ Q_i P_j - Q_j P_i\}\, .
\ee
Using \refb{e8} and \refb{egcd} we now see that
for the charge vector \refb{e6} we have
\be \label{e8b}
\psi(q) = \gcd\{ n, Q_1Q_5, J, Q_5\} 
= \gcd \{ n, Q_5\}\, ,
\ee
since $Q_1$ and $J$ are divisible by
$Q_5$. The condition \refb{e1} now gives
\be \label{e8c}
\gcd\{ n, Q_5\} = 1\, .
\ee

Our goal is to prove \refb{e5} for the charge vector given in
\refb{e6}.  A 
formula for the index $B_{14}$ for unit torsion 
dyons has been written down in 
\cite{0506151,0506228,0803.1014}, 
but since the charge
vector \refb{e6} has torsion $Q_5$, we cannot use the results
of \cite{0506151,0506228,0803.1014} 
directly. We shall now show however that by a
suitable U-duality transformation we can map the charge
vector \refb{e6} to another charge vector of unit torsion, and then
use the result of \cite{0803.1014} to compute $d(q)$. For this
we recall that the U-duality group of type IIA string theory
compactified 
on $T^4$ contains a string-string duality transformation that
exchanges the fundamental string with an NS 5-brane wrapped
on $T^4$. Applying this duality transformation on the charge
vector \refb{e6} exchanges the quantum number $n$ representing
fundamental string winding charge along $S^1$ with the
quantum number $Q_1$ representing NS 5-brane winding charge
along $T^4\times S^1$. Thus the new charge vectors $Q'$ and $P'$
are of the form
\be \label{e8d}
Q'= \pmatrix{0 \cr Q_1\cr 0\cr 1}, 
\quad P'=\pmatrix{n\cr J\cr Q_5\cr 0}\, ,
\quad n,Q_1,Q_5,J\in\ZZZ, \quad Q_5|J, Q_1\, .
\ee
For this we have
\be \label{e8e}
Q^{\prime 2} = 2 Q_1, \quad P^{\prime 2} = 2 n Q_5,
\quad Q'\cdot P' = J\, ,
\ee
and
\be \label{e8f}
\gcd \{Q'\wedge P'\} = \gcd(n, J, Q_5) = 1\, ,
\ee
using \refb{e8c}. Due to \refb{e8f} the new
charge vectors $(Q',P')$ belong to the class for which the
degeneracy of states was analyzed in \cite{0803.1014}. The
result for the degeneracy may be stated as
\be \label{e8g}
d'(q) = \sum_{s|Q^{\prime 2}/2, P^{\prime 2}/2,Q'\cdot P'}
s\, \wh c\left((Q^{\prime 2} P^{\prime 2}-(Q'\cdot P')^2)/s^2\right)
= \sum_{s|Q_1, n Q_5, J} s\, \wh c ((4 n Q_1 Q_5 - J^2)/s^2)\, .
\ee

Our task now is to show that \refb{e5} reduces to 
$d'(q)$ given in \refb{e8g}
for the choice of charge vectors given in \refb{e6},
or equivalently \refb{e8d}. For definiteness we shall work with
the charge vector given in \refb{e6}. First of all we note from
\refb{e3} that here $\Delta(q)=4 n Q_1 Q_5 - J^2$. Thus
\refb{e5} takes the form
\be \label{e8h}
d(q) = \sum_{2s|\chi(q)} 
s\, \wh c ((4 n Q_1 Q_5 - J^2)/s^2)\, .
\ee
We shall now analyze the condition on $s$ imposed by the
requirement $2s|\chi(q)$. For this we first need to construct
the vector $\wt q_a$ appearing in the expression \refb{e4}
of $\chi(q)$. Now $\wt q_a$ is a trilinear in $q_a$ transforming
in the ${\bf 56}$ representation of $E_{7(7)}(\ZZZ)$. Using the
fact that {\bf 56} decomposes into a {\bf (2,12)} and a {\bf (1,32)}
representation of $SL(2,\ZZZ)\times O(6,6;\ZZZ)$ and that we
have set the RR components of $q$ to zero, we see that the only
non-zero component of $\wt q_a$ must be the {\bf (2,12)} components
constructed from $Q_i$ and $P_i$. Up to a normalization the
unique trilinear combination of 
$Q_i$'s and $P_i$'s transforming in the
{\bf (2,12)} representation is
\be \label{e8k}
\pmatrix{\wt Q\cr \wt P} 
= \pmatrix{Q^2 P - (Q\cdot P) Q\cr -P^2 Q + (Q\cdot P) P}\, .
\ee
This is in fact also the correct normalization since with this choice
$(\wt Q, \wt P)$ satisfies the constraint given in \refb{e2}, \refb{e3}:
\be \label{e8m}
{1\over 2}\LLL_{ab}\wt q_a q_b = {1\over 2}
\epsilon_{\alpha\beta} L_{ij} \wt M_{i\alpha} 
M_{j\beta}=
{1\over 2} \left(\wt  Q\cdot \  P - \wt P \cdot Q\right)  = 
Q^2 P^2 - (Q\cdot P)^2\, .
\ee
Using \refb{e6} and \refb{e8k} we get
\be \label{esk1}
\wt Q = \pmatrix{2nQ_1 \cr nJ \cr 2nQ_5\cr -J}, 
\qquad \wt P = \pmatrix{JQ_1\cr J^2 - 2 Q_1 Q_5 n\cr
JQ_5\cr -2Q_1Q_5}\, ,
\ee
and hence
\ben \label{esk2}
\chi(q) &=& \gcd\{q_a \wt q_b - \wt q_a q_b\}
= \gcd\{ Q_i \wt Q_j - Q_j \wt Q_i, P_i \wt P_j - P_j \wt P_i,
Q_i \wt P_j - P_j \wt Q_i\}\nonumber \\
&=&  \gcd\{2nQ_1,2nJ,2nQ_5, 2JQ_1,2JQ_5,2Q_1Q_5,2J^2\}\, .
\een
In the last line of \refb{esk2} we have dropped terms which contain
one of the terms appearing inside $\{~\}$ as factors.
We now use the fact that $Q_5$ divides $J$ and $Q_1$. In this case
$2Q_5$ is a factor of every term inside $\{~\}$ in the right hand side
of \refb{esk2} and we can write
\be \label{esk3}
\chi(q) = 2 Q_5 \, \gcd\{n, Q_1,J \}\, .
\ee
For configurations with $\psi(q)=1$, \refb{e8c} 
gives $\gcd\{n,Q_5\}=1$. As a result we can drop factors of $Q_5$
from the $Q_1$ and $J$ term inside $\{~\}$ in \refb{esk3}
and express this equation as
\be \label{esk4}
\chi(q) = 2 Q_5 \, \gcd\{n, Q_1/Q_5, J/Q_5\} =
2 \gcd\{ nQ_5, Q_1, J\}\, .
\ee
Using this we may express \refb{e8h} as
\be \label{sk5}
d(q) = \sum_{s|nQ_5,Q_1,J} 
s\, \wh c ((4 n Q_1 Q_5 - J^2)/s^2)\, .
\ee
This in in precise agreement with \refb{e8g}, thereby proving
\refb{e5}.

It will be interesting to try to relax the constraint $\psi(q)=1$
on the dyon charges. One could try to guess the answer following
the approach of \cite{0802.1556} or 
try to
determine the spectrum directly
by analyzing the D1-D5-KK monopole system along the line of
\cite{0803.2692}.

\medskip

\noindent {\bf Acknowledgment:} I would like to thank
Shamik Banerjee
for useful discussions. I would also like to acknowledge
the hospitality of the Abdus Salam
International Centre for Theoretical Physics where part of this
work was done. Finally, I would like to thank the people of
India for supporting research in string theory.


\end{document}